\documentclass[fleqn,usenatbib]{mnras}

\usepackage{newtxtext,newtxmath}

\usepackage[T1]{fontenc}

\DeclareRobustCommand{\VAN}[3]{#2}
\let\VANthebibliography\thebibliography
\def\thebibliography{\DeclareRobustCommand{\VAN}[3]{##3}\VANthebibliography}

\usepackage{graphicx}	
\usepackage{amsmath}	

\title[Host galaxy magnitude of OJ~287 from its colours at minimum light]{Host galaxy magnitude of OJ~287 from its colours at minimum light}

\author[Valtonen et al.]{Mauri~J.~Valtonen,$^{1,2}$\thanks{E-mail: mvaltonen2001@yahoo.com}
Lankeswar~Dey,$^{3}$
S.~Zola,$^{4,5}$
S.~Ciprini,$^{6,7}$
M.~Kidger,$^{8}$
T.~Pursimo,$^{9}$
\newauthor A.~Gopakumar,$^{3}$
K.~Matsumoto,$^{10}$
K.~Sadakane,$^{10}$
D.~B.~Caton,$^{11}$
K.~Nilsson,$^{1}$
S.~Komossa,$^{12}$
\newauthor M.~Bagaglia,$^{13}$
A.~Baransky,$^{14}$
P.~Boumis,$^{15}$
D.~Boyd,$^{16}$
A.~J.~Castro-Tirado,$^{17}$
B.~Debski,$^{4}$
\newauthor M.~Drozdz,$^{5}$
A.~Escartin P{\'e}rez,$^{18}$
M.~Fiorucci,$^{13}$
F.~Garcia,$^{19}$
K.~Gazeas,$^{20}$
S.~Ghosh,$^{3}$
\newauthor V.~Godunova,$^{21}$
J.~L.~Gomez,$^{17}$
R.~Gredel,$^{22}$
D.~Grupe$^{23}$
J.~B.~Haislip,$^{24}$
T.~Henning,$^{22}$
G.~Hurst,$^{25}$
\newauthor J.~Jan{\'i}k,$^{26}$
V.~V.~Kouprianov,$^{24}$
H.~Lehto,$^{2}$
A.~Liakos,$^{15}$
S.~Mathur,$^{27,28}$
M.~Mugrauer,$^{29}$
\newauthor R.~Naves Nogues,$^{30}$
G.~Nucciarelli$^{13}$
W.~Ogloza,$^{5}$
D.~K.~Ojha,$^{3}$
U.~Pajdosz-{\'S}mierciak,$^{4}$
S.~Pascolini,$^{13}$
\newauthor G.~Poyner,$^{31}$
D.~E.~Reichart,$^{24}$
N.~Rizzi,$^{13}$
F.~Roncella,$^{13}$
D.~K.~Sahu,$^{32}$
A.~Sillanp\"a\"a,$^{2}$
A.~Simon,$^{33}$
M.~Siwak,$^{5,34}$
\newauthor F.~C.~ Sold{\'a}n Alfaro,$^{35,36}$
E.~Sonbas,$^{37,38}$
G.~Tosti,$^{13}$
V.~Vasylenko,$^{33}$
J.~R.~Webb,$^{39}$
and P.~Zielinski,$^{40}$
\\
$^{1}$Finnish Centre for Astronomy with ESO, University of Turku, Finland\\
$^{2}$Tuorla Observatory, Department of Physics and Astronomy, University of Turku, Finland\\
$^{3}$Department of Astronomy and Astrophysics, Tata Institute of Fundamental Research, Mumbai 400005, India\\
$^{4}$Astronomical Observatory, Jagiellonian University, ul. Orla 171, PL-30-244 Krakow, Poland\\
$^{5}$Mt Suhora Observatory, Pedagogical University, ul. Podchorazych 2, PL-30-084 Krakow, Poland\\
$^{6}$Space Science Data Center - Agenzia Spaziale Italiana, via del Politecnico, snc, I-00133, Roma, Italy\\
$^{7}$Instituto Nazionale di Fisica Nucleare, Sezione di Perugia, Perugia I-06123, Italy\\
$^{8}$European Space Agency European Space Astronomy Centre, 28691 Villanueva de la Ca{\~n}ada, Madrid, Spain\\
$^{9}$Nordic Optical Telescope, Apartado 474, E-38700 Santa Cruz de La Palma, Spain\\
$^{10}$Astronomical Institute, Osaka Kyoiku University, 4-698 Asahigaoka, Kashiwara, Osaka 582-8582, Japan\\
$^{11}$Dark Sky Observatory, Department of Physics and Astronomy, Appalachian State University, Boone, NC 28608, USA\\
$^{12}$Max-Planck-Institut f\"ur Radioastronomie, Auf dem H{\"u}gel 69, D-53121 Bonn, Germany\\
$^{13}$Physics Department, University of Perugia, via A. Pascoli, 06123, Perugia, Italy \\
$^{14}$Astronomical Observatory of Taras Shevshenko National University of Kyiv, Observatorna str. 3, 04053 Kyiv, Ukraine\\
$^{15}$Institute for Astronomy, Astrophysics, Space Applications and Remote Sensing, National Observatory of Athens, \\ Metaxa \& Vas. Pavlou St., Penteli, Athens GR-15236, Greece\\
$^{16}$BAA Variable Star Section, West Challow Observatory, OX12 9TX, UK\\
$^{17}$Instituto de Astrof\'isica de Andaluc\'ia (IAA-CSIC), P.O. Box 03004, E-18080, Granada, Spain\\
$^{18}$Aritz Bidea No 8 4B (48100) Mungia Bizkaia, Spain\\
$^{19}$Mu\~nas de Arriba La Vara, Vald{\'e}s (MPC J38) 33780 Vald\'es, Asturias -- Spain\\
$^{20}$Section of Astrophysics, Astronomy and Mechanics, Department of Physics, National and Kapodistrian University of Athens, GR-15784 
Zografos, Athens, Greece\\
$^{21}$ICAMER Observatory of NASU, 27, Acad. Zabolotnoho str., 03143 Kyiv, Ukraine\\
$^{22}$Max Planck Institute for Astronomy, Koenigstuhl 17, D-69117 Heidelberg, Germany\\
$^{23}$Dept. of Physics, Earth Science, and Space System Engineering, Morehead State University, 235 Martindale Dr, Morehead, KY 40351, USA\\
$^{24}$University of North Carolina at Chapel Hill, Chapel Hill, North Carolina NC 27599, USA\\
$^{25}$16 Westminster Close Basingstoke Hampshire RG22 4PP, UK\\
$^{26}$Department of Theoretical Physics and Astrophysics, Masaryk University, Kotlarska 2, 61137 Brno, Czech Republic\\
$^{27}$Department of Astronomy, The Ohio State University, 140 West 18th Avenue, Columbus, OH 43210, USA\\
$^{28}$Center for Cosmology and Astroparticle Physics, The Ohio State University, 191 West Woodruff Avenue, Columbus, OH 43210, USA\\
$^{29}$Astrophysikalisches Institut und Universitäts-Sternwarte, Schillergässchen 2-3, D-07745 Jena, Germany\\
$^{30}$C/Jaume Balmes No 24, Cabrils, Barcelona E-08348, Spain\\
$^{31}$BAA Variable Star Section, 67 Ellerton Road, Kingstanding, Birmingham B44 0QE, UK\\
$^{32}$Indian Institute of Astrophysics, II Block, Koramangala, Bengaluru 560 034, India\\
$^{33}$Astronomy and Space Physics Department, Taras Shevshenko National University of Kyiv, Volodymyrska str. 60, 01033 Kyiv, Ukraine\\
$^{34}$Konkoly Observatory, Research Centre for Astronomy and Earth Sciences, Eötvös Loránd Research Network (ELKH), Hungarian Academy of Sciences, Konkoly-Thege Mikl\'os \'ut 15--17, 1121 Budapest, Hungary\\
$^{35}$C/Petrarca 6 1{$^a$} 41006 Sevilla, Spain\\
$^{36}$Astrophysics and Astronomy Introduction, Expert Teacher, Experience Seminar, Seville University, Seville, Spain\\
$^{37}$Department of Physics, University of Adiyaman, Adiyaman 02040, Turkey\\
$^{38}$Astrophysics Application and Research Center, Adiyaman University, Adiyaman 02040, Turkey\\
$^{39}$Florida International University and SARA Observatory, University Park Campus, Miami, FL 33199, USA\\
$^{40}$Institute of Astronomy, Faculty of Physics, Astronomy and Informatics, Nicolaus Copernicus University in Toru{\'n}, ul. Grudzi\k{a}dzka 5, 87-100 Toru{\'n}, Poland\\
}

\date{Accepted XXX. Received YYY; in original form ZZZ}

\pubyear{2015}

\begin{document}
\label{firstpage}
\pagerange{\pageref{firstpage}--\pageref{lastpage}}
\maketitle

\begin{abstract}
OJ~287 is a BL Lacertae type quasar in which the active galactic nucleus (AGN) outshines the host galaxy by an order of magnitude. 
The only exception to this may be at minimum light when the AGN activity is so low that the host galaxy may make quite a considerable contribution to the photometric intensity of the source. 
Such a dip or a fade in the intensity of OJ~287 occurred in November 2017, when its brightness was about 1.75 magnitudes lower than the recent mean level. 
We compare the observations of this fade with similar fades in OJ~287 observed earlier in 1989, 1999, and 2010.
It appears that there is a relatively strong reddening of the B$-$V colours of OJ~287 when its V-band brightness drops below magnitude 17.
Similar changes are also seen V$-$R, V$-$I, and R$-$I colours during these deep fades.
These data support the conclusion that the total magnitude of the host galaxy is $V=18.0 \pm 0.3$, corresponding to $M_{K}=-26.5 \pm 0.3$ in the K-band.
This is in agreement with the results, obtained using the integrated surface brightness method, from recent surface photometry of the host.
These results should encourage us to use the colour separation method also in other host galaxies with strongly variable AGN nuclei. 
In the case of OJ~287, both the host galaxy and its central black hole are among the biggest known, and its position in the black hole mass-galaxy mass diagram lies close to the mean correlation.
\end{abstract}




\section{Introduction}
\label{sec:intro}

OJ~287 is a BL Lacertae type quasar (RA: 08:54:48.87, Dec: +20:06:30.6), situated at a redshift of $z = 0.306$. 
Based on the faintness of its optical Balmer lines from the broad-line region \citep{sit85, nil10}, OJ~287 is classified as a BL Lac object.  
Given its spectral-energy distribution (SED), it is of LBL type \citep{Padovani1995}. Optical emission lines from the narrow-line region (NLR) like [OIII]$\lambda$5007 are only faintly present \citep{nil10}. 
It shows high brightness activity at about 12 yr intervals, most likely due to a binary black hole (BH) central engine having that orbital period \citep{sil88, dey19}. 
The binary model consists of two BHs orbiting each other in an orbit of eccentricity $\sim 0.65$ and semi-major axis $\sim 0.05$ pc \citep{dey18}. 
The components are unequal: the mass ratio is $\sim 120$. 
The secondary cycle of activity with $\sim$ 56-year period is also well established and it is understood as the precession cycle of the binary \citep{val10}.

\begin{figure}
\includegraphics[width=0.95 \columnwidth] {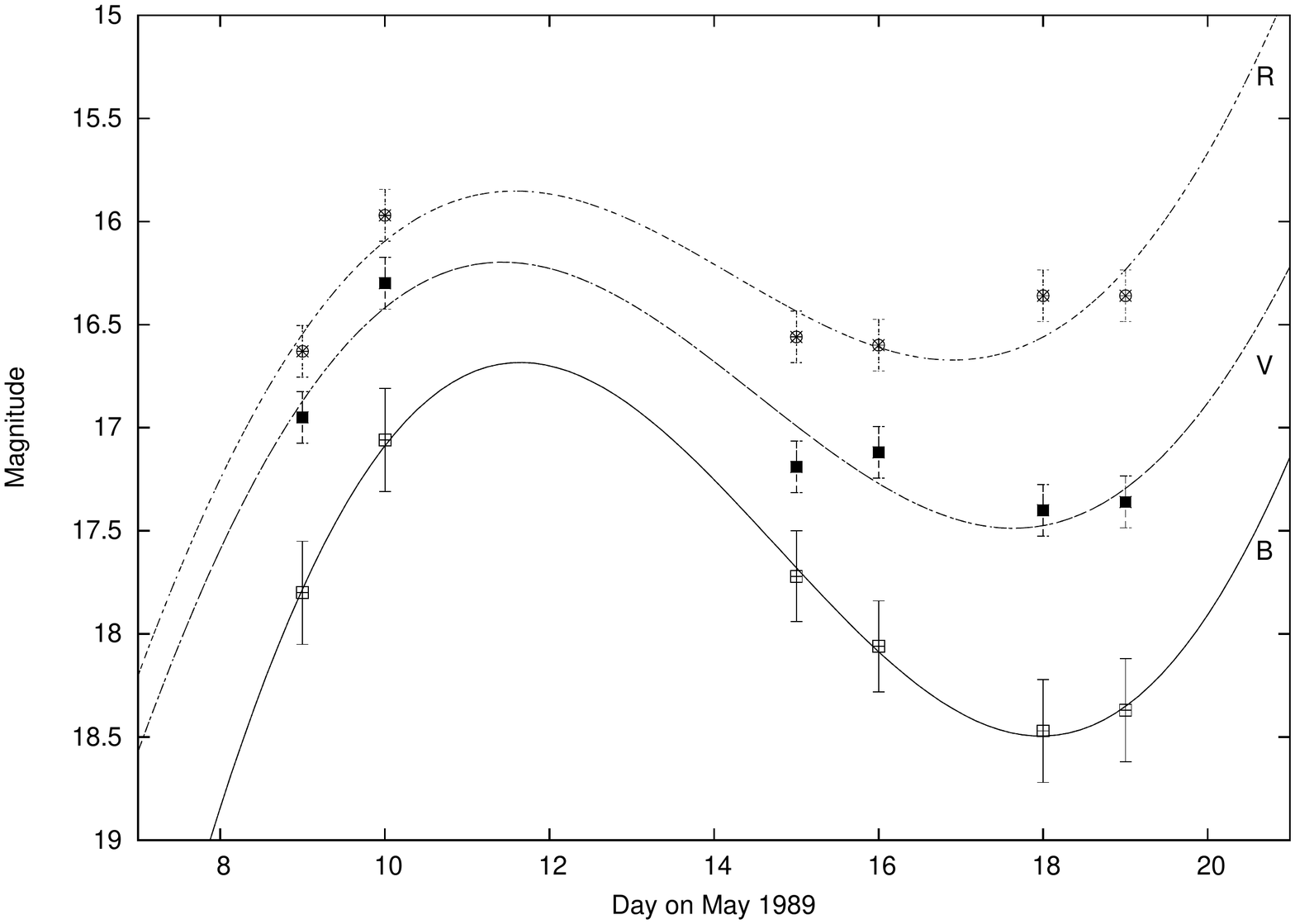}
\caption{The observations of the 1989 fade of OJ~287 with the {\it Nordic Optical Telescope} and the {\it Jacobus Kapteyn Telescope} at La Palma, with $2 \sigma$ errorbars. Here we plot the magnitudes in three filters B, V and R. The lines are fifth order polynomial fits through the data points.}
\label{fig:1989fade}
\end{figure}

It is less well known that OJ~287 also shows prominent low activity states called fades, at similar intervals \citep{tak90,pie99}.
First such prominent dip in the brightness of OJ~287 was noticed during 1989.
\citet{tak90} carried out a detailed study of the 1989 fade and concluded that at the minimum light, the host galaxy radiated at least the same amount of energy as the synchrotron source in the V-band.
The evidence for seeing the host galaxy comes in the colours of the object: while the object's brightness fades, its spectral energy distribution changes from the usual power-law spectrum in the optical region to the spectrum typical of an elliptical galaxy at the redshift of z = 0.3 of OJ~287. In Figure~\ref{fig:1989fade}, we show the multicolour data of OJ~287 during the 1989 fade.

\cite{leh96} predicted that the fade should be repeated in 1998, one binary orbital cycle later, when the underlying binary black hole system is at the same orbital phase as in 1989. 
Due to a strong forward precession of the major axis of the orbit, the orbital phase covers the whole cycle in less than 12 years. 
The predicted fade indeed took place, only 6 weeks after the predicted time \citep{pie99}. 
Since then, there have been prominent fades in 1999 and 2010, while the most recent one took place in November 2017.

In this paper, we collect the data from the fades since 1989 and compare them with the \citet{tak90} data.
We make use of these deepest fades to try to detect the underlying host galaxy. 
The host has been difficult to detect because the radiation from the active galactic nucleus (AGN) is strongly beamed toward us. 
During a fade, we have the best opportunity to see the host galaxy.
Further, it is important to determine the host brightness in order to get another point at the upper end of the BH mass vs. host bulge stellar mass diagram, where the data are few \citep{sag16}.

The organisation of the paper is as follows: In section~\ref{sec:observations}, we describe the photometric observations of different fades of OJ~287 used in this paper. 
Section~\ref{sec:OJ_colours} uses these and earlier photometric observations to calculate the changes of colour of OJ~287 as a function of its magnitude. 
The changes are modelled by a combination of light from the AGN and the host galaxy. 
In this way the two contributions are separated, and the host magnitude is derived. 
In section~\ref{sec:discussion}, we compare this determination of the host magnitude with the integrated surface photometry method, and find good agreement with a recent K-band determination of the host magnitude.

\begin{figure*}
\includegraphics[width=0.48 \textwidth] {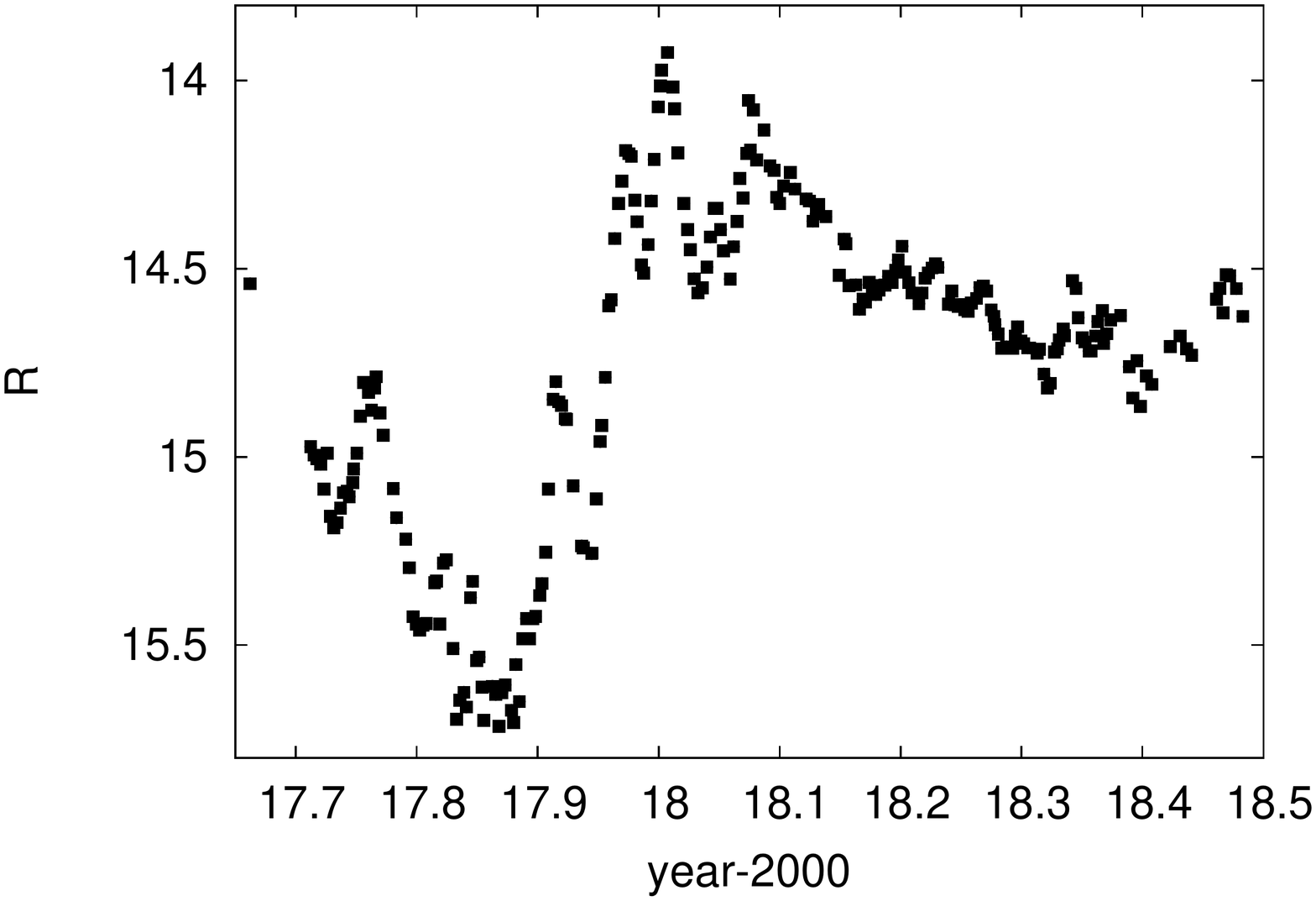}
\includegraphics[width=0.48 \textwidth] {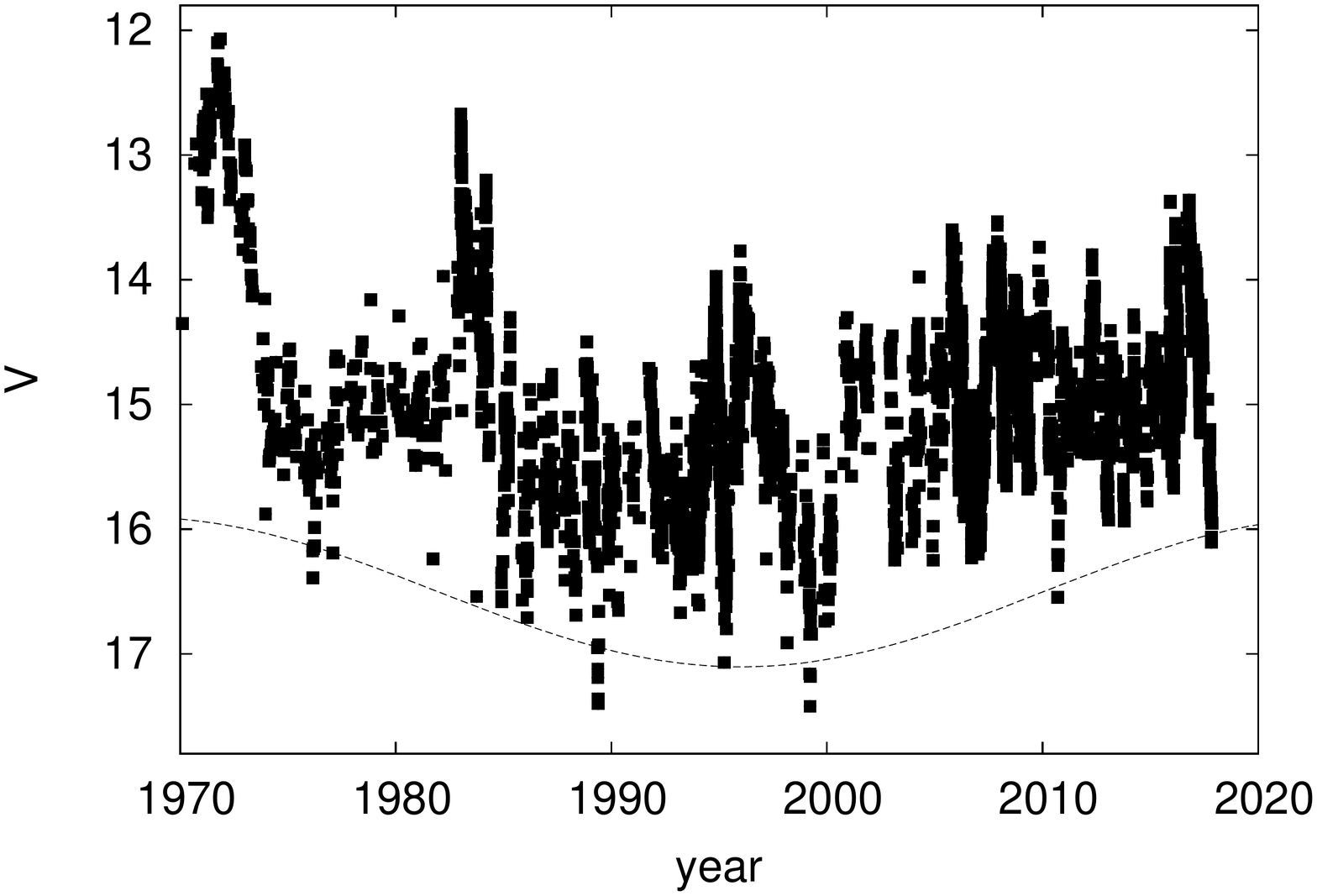}

\caption{Left panel: The R-band optical light curve of OJ~287 in the 2017/18 season. Right panel: The V-band optical light curve of OJ~287 from 1970 to 2018. The dotted line runs 1.75 magnitudes below the historical mean brightness of OJ~287.}
\label{fig:2017_and_historical_fades}
\end{figure*}

\section{Observations}
\label{sec:observations}
In anticipation of the big flare in OJ~287 on December 2015, an observational campaign was set up \citep{val16}. 
After the detection of the flare which peaked on December 5, we continued to monitor the blazar in the wide band R filter with the Skynet Robotic Network telescopes and at sites in Japan, Greece, Ukraine, UK, Spain, and Poland. 
We achieved a daily cadence except for the periods when OJ~287 was too close to the sun or the moon. 
On several occasions, we also took multi-filter data. 
All photometric observations were reduced by a single person (S.Z.) to provide as uniform as possible data set.  
The reduction is done in a standard way: images were calibrated for bias, dark and flatfield and we extracted magnitudes using the aperture method. 
More details about the procedure were given in \citet{pih13}.

The left panel of Figure~\ref{fig:2017_and_historical_fades} shows OJ~287 observations which we gathered during the 2017/18 season.
In the right panel of Figure~\ref{fig:2017_and_historical_fades}, we present how this recent fade compares with earlier ones. 
We note that the minimum light at 2017.85 is 1.75 magnitudes below the oscillating historical average shown by the dotted line (oscillation period is $\sim$ 56 yr \citep{val06}).

\begin{table*}
\centering
\caption{Perugia VRI observations of the 1998 and 1999 fades.}
\label{tab:dataPer98}
\begin{tabular}{cccc cccc}

\hline
$Year$  &    $V$          &  $R$  & $I$ &  $Year$  &    $V$          &  $R$  & $I$                          \\
\hline

1998.0218&  16.08 $\pm$ 0.11&  15.26 $\pm$ 0.05&  14.56 $\pm$ 0.03  &1998.9614 & 16.35 $\pm$ 0.19& 15.85 $\pm$ 0.07&      15.15 $\pm$  0.05    \\

1998.0223&  16.09 $\pm$ 0.09&  15.57 $\pm$ 0.06&  14.71 $\pm$ 0.04  &1998.9616 &              ---& 15.71 $\pm$ 0.07&      15.10 $\pm$  0.064  \\

1998.1312&  16.40 $\pm$ 0.13&  16.09 $\pm$ 0.08&  14.96 $\pm$ 0.04  &1998.9637 & 15.83 $\pm$ 0.18& 15.89 $\pm$ 0.06&      15.10 $\pm$  0.045  \\

1998.1338&  15.92 $\pm$ 0.06&  15.72 $\pm$ 0.04&  15.27 $\pm$ 0.04  &1998.9642 &              ---& 15.68 $\pm$ 0.06&      15.17 $\pm$  0.05  \\

1998.1366&  16.91 $\pm$ 0.13&  16.01 $\pm$ 0.06&  15.10 $\pm$ 0.04  &1998.9748 &              ---& 15.77 $\pm$ 0.05&      14.95 $\pm$  0.04  \\

1998.1394&  16.53 $\pm$ 0.13&  16.34 $\pm$ 0.10&  15.01 $\pm$ 0.04  &1999.0128 & 16.26 $\pm$ 0.23& 15.51 $\pm$ 0.05&      14.87 $\pm$  0.04  \\

1998.1421&              ---&  15.69 $\pm$ 0.07&  15.08 $\pm$ 0.04  &1999.0134 &              ---& 15.68 $\pm$ 0.06&      14.83 $\pm$  0.04  \\

1998.1559&  16.78 $\pm$ 0.14&  15.79 $\pm$ 0.06&  15.15 $\pm$ 0.04  &1999.0136 &              ---& 15.63 $\pm$ 0.06&      14.86 $\pm$  0.04  \\

1998.1586&  16.63 $\pm$ 0.12&  15.67 $\pm$ 0.05&  15.17 $\pm$ 0.04  &1999.0157 & 15.88 $\pm$ 0.19& 15.67 $\pm$ 0.05&      15.07 $\pm$  0.04  \\

1998.1668&  16.41 $\pm$ 0.20&  15.61 $\pm$ 0.08&  15.00 $\pm$ 0.05  &1999.0164 &              ---& 15.76 $\pm$ 0.13&      14.92 $\pm$  0.08  \\

1998.1695&  15.94 $\pm$ 0.10&  15.59 $\pm$ 0.06&  15.03 $\pm$ 0.05  &1999.0239 &              ---& 15.92 $\pm$ 0.07&      15.02 $\pm$  0.04 \\

1998.1777&  15.96 $\pm$ 0.06&  15.40 $\pm$ 0.05&      ---          &1999.0951&              ---& 15.67 $\pm$ 0.09&      15.25 $\pm$  0.08 \\

1998.1887&              ---&  15.57 $\pm$ 0.06&  14.86 $\pm$ 0.04  &1999.1058&              ---& 15.82 $\pm$ 0.07&      15.07 $\pm$  0.05  \\

1998.1996&  16.30 $\pm$ 0.11&  15.78 $\pm$ 0.06&  14.96 $\pm$ 0.04  &1999.1196&              ---& 15.44 $\pm$ 0.06&      14.78 $\pm$  0.04  \\

1998.2049&  15.72 $\pm$ 0.10&  15.47 $\pm$ 0.06&  14.48 $\pm$ 0.04  &1999.1524&              ---& 16.00 $\pm$ 0.07&      15.34 $\pm$  0.05  \\

1998.2103&  15.82 $\pm$ 0.08&  15.33 $\pm$ 0.05&  14.75 $\pm$ 0.04  &1999.1551&              ---& 15.53 $\pm$ 0.06&      15.09 $\pm$  0.05  \\

1998.2158&              ---&  15.40 $\pm$ 0.12&  14.76 $\pm$ 0.10  &1999.1688&              ---& 16.13 $\pm$ 0.12&      15.56 $\pm$  0.07  \\

1998.2350&              ---&  15.48 $\pm$ 0.05&  14.85 $\pm$ 0.04  &1999.1825&              ---& 16.19 $\pm$ 0.10&      15.56 $\pm$  0.06  \\

1998.2377&  16.51 $\pm$ 0.16&  15.63 $\pm$ 0.06&  14.92 $\pm$ 0.05  &1999.1880&              ---& 16.22 $\pm$ 0.22&      15.80 $\pm$  0.16  \\

1998.2405&  16.35 $\pm$ 0.21&  15.82 $\pm$ 0.08&        ---        &1999.1907&              ---& 16.23 $\pm$ 0.09&      15.68 $\pm$  0.07 \\

1998.2432&  16.07 $\pm$ 0.20&                ---&  14.90 $\pm$ 0.05  &1999.1934&              ---& 16.02 $\pm$ 0.08&      15.41 $\pm$  0.06  \\

1998.2461&  16.05 $\pm$ 0.19&  15.65 $\pm$ 0.13&  14.81 $\pm$ 0.07  &1999.1961&              ---& 15.88 $\pm$ 0.08&      15.16 $\pm$  0.05  \\

1998.2541&  15.90 $\pm$ 0.11&  15.41 $\pm$ 0.06&  14.77 $\pm$ 0.05  &1999.2016&              ---& 16.02 $\pm$ 0.07&      15.62 $\pm$  0.06  \\

1998.2568&  16.01 $\pm$ 0.16&  15.45 $\pm$ 0.06&  14.59 $\pm$ 0.04  &1999.2044&              ---& 16.12 $\pm$ 0.11&      15.37 $\pm$  0.06  \\

1998.2705&  15.79 $\pm$ 0.09&  15.44 $\pm$ 0.05&  14.84 $\pm$ 0.04  &1999.2153&              ---& 17.02 $\pm$ 0.18&      15.85 $\pm$  0.10  \\

1998.3006&  15.96 $\pm$ 0.08&  15.29 $\pm$ 0.05&          ---      &1999.2155&              ---& 16.31 $\pm$ 0.10&      15.69 $\pm$  0.07 \\

1998.3034&  16.00 $\pm$ 0.09&  15.41 $\pm$ 0.05&  14.75 $\pm$ 0.04  &1999.2180&              ---& 15.94 $\pm$ 0.08&      15.44 $\pm$  0.06  \\

1998.3061&  15.89 $\pm$ 0.13&  15.24 $\pm$ 0.05&  14.78 $\pm$ 0.05  &1999.2235&              ---& 16.78 $\pm$ 0.15&      15.95 $\pm$  0.08  \\

1998.3089&  15.86 $\pm$ 0.09&  15.42 $\pm$ 0.07&  14.74 $\pm$ 0.04  &1999.2453&              ---& 16.34 $\pm$ 0.15&      16.09 $\pm$  0.11  \\

1998.3144&  15.39 $\pm$ 0.09&  15.24 $\pm$ 0.07&  14.72 $\pm$ 0.05  &1999.2508&              ---& 16.24 $\pm$ 0.12&      15.73 $\pm$  0.09  \\

1998.3171&              ---&  15.20 $\pm$ 0.09&  14.58 $\pm$ 0.06  &1999.2562&              ---& 16.21 $\pm$ 0.11&      15.94 $\pm$  0.09  \\

1998.9583 & 16.89 $\pm$ 0.38&  16.01 $\pm$ 0.09&  14.85 $\pm$ 0.04  &1999.2727&              ---& 16.44 $\pm$ 0.11&      15.84 $\pm$  0.08    \\

1998.9610 &              ---&  15.76 $\pm$ 0.05&  15.07 $\pm$  0.04 &        &                  &                & \\

\hline

\end{tabular}
\end{table*}

\begin{table*}
\begin{center}
\caption{Krakow and Mt.~Suhora BVRI observations of the 2006, 2010 and 2017 fades.}
\label{tab:dataKRKSUH06}
\vskip 0.1cm
\begin{tabular}{lcccc}
\hline
$Year$&        $B$&              $V$&              $R$    &      $I$  \\
\hline

2006.7400&      ---      & 15.85 $\pm$ 0.03& 15.42 $\pm$ 0.02& 14.77 $\pm$ 0.01    \\
2006.7455&      ---      & 15.72 $\pm$ 0.02& 15.30 $\pm$ 0.02& 14.66 $\pm$ 0.01    \\
2006.7482&      ---      & 15.64 $\pm$ 0.01& 15.21 $\pm$ 0.01& 14.56 $\pm$ 0.01    \\
2006.7647&      ---      & 15.54 $\pm$ 0.02& 15.11 $\pm$ 0.01& 14.46 $\pm$ 0.01    \\
2006.7730&      ---      & 15.44 $\pm$ 0.01& 15.01 $\pm$ 0.01& 14.36 $\pm$ 0.01    \\
2006.7756&      ---      & 15.60 $\pm$ 0.01& 15.15 $\pm$ 0.01& 14.50 $\pm$ 0.01    \\
2006.7757&      ---      &      ---      & 15.12 $\pm$ 0.03& 14.43 $\pm$ 0.02    \\
2006.7783&      ---      &      ---      & 15.09 $\pm$ 0.01& 14.45 $\pm$ 0.02    \\
2006.7921&      ---      & 15.85 $\pm$ 0.02& 15.40 $\pm$ 0.01& 14.74 $\pm$ 0.02    \\
2006.7948&      ---      & 16.01 $\pm$ 0.02& 15.57 $\pm$ 0.01& 14.89 $\pm$ 0.01    \\
2006.7940&16.61 $\pm$ 0.04& 16.02 $\pm$ 0.02& 15.57 $\pm$ 0.01& 14.85 $\pm$ 0.02    \\
2006.7976&      ---      & 15.97 $\pm$ 0.04& 15.50 $\pm$ 0.01& 14.84 $\pm$ 0.02    \\
2006.8003&      ---      & 15.68 $\pm$ 0.02& 15.26 $\pm$ 0.02& 14.62 $\pm$ 0.01    \\
2006.8524&16.56 $\pm$ 0.03& 16.01 $\pm$ 0.03& 15.56 $\pm$ 0.02& 14.92 $\pm$ 0.02    \\
2006.8551&16.23 $\pm$ 0.08& 15.60 $\pm$ 0.11& 15.18 $\pm$ 0.10& 14.57 $\pm$ 0.08    \\
2006.8742&16.14 $\pm$ 0.04& 15.59 $\pm$ 0.02& 15.17 $\pm$ 0.02& 14.54 $\pm$ 0.02    \\
2006.8800&16.16 $\pm$ 0.03& 15.59 $\pm$ 0.01& 15.17 $\pm$ 0.01& 14.52 $\pm$ 0.01    \\
2006.9070&15.97 $\pm$ 0.01& 15.43 $\pm$ 0.02& 15.02 $\pm$ 0.01& 14.38 $\pm$ 0.01    \\
2006.9097&16.06 $\pm$ 0.01& 15.49 $\pm$ 0.01& 15.07 $\pm$ 0.01& 14.44 $\pm$ 0.01    \\
2006.9150&16.09 $\pm$ 0.02& 15.51 $\pm$ 0.02& 15.09 $\pm$ 0.02& 14.44 $\pm$ 0.02    \\
2006.9234&16.13 $\pm$ 0.05& 15.56 $\pm$ 0.01& 15.10 $\pm$ 0.02& 14.46 $\pm$ 0.02    \\
2006.9290&16.08 $\pm$ 0.01& 15.53 $\pm$ 0.02& 15.11 $\pm$ 0.03& 14.47 $\pm$ 0.03    \\
2006.9452&16.41 $\pm$ 0.03& 15.82 $\pm$ 0.02& 15.35 $\pm$ 0.02& 14.70 $\pm$ 0.01    \\
2006.9534&16.36 $\pm$ 0.01& 15.78 $\pm$ 0.01& 15.32 $\pm$ 0.01& 14.65 $\pm$ 0.02    \\
2006.9588&16.48 $\pm$ 0.01& 15.88 $\pm$ 0.02& 15.45 $\pm$ 0.02& 14.78 $\pm$ 0.02    \\
2006.9863&16.32 $\pm$ 0.02& 15.74 $\pm$ 0.02& 15.31 $\pm$ 0.01& 14.65 $\pm$ 0.02    \\
2006.9889&16.37 $\pm$ 0.01& 15.80 $\pm$ 0.01& 15.36 $\pm$ 0.01& 14.69 $\pm$ 0.01    \\
2006.9946&16.55 $\pm$ 0.02& 15.98 $\pm$ 0.02& 15.54 $\pm$ 0.01& 14.88 $\pm$ 0.01    \\
2007.0270&16.43 $\pm$ 0.04& 15.85 $\pm$ 0.02& 15.39 $\pm$ 0.02& 14.72 $\pm$ 0.02    \\
2007.0328&      ---      & 15.77 $\pm$ 0.06& 15.33 $\pm$ 0.03& 14.69 $\pm$ 0.03    \\
2007.0386&16.62 $\pm$ 0.02& 16.03 $\pm$ 0.02& 15.57 $\pm$ 0.01& 14.89 $\pm$ 0.02    \\
2007.0442&16.60 $\pm$ 0.03& 16.00 $\pm$ 0.02& 15.55 $\pm$ 0.03& 14.94 $\pm$ 0.02    \\
2007.0708&      ---      & 16.05 $\pm$ 0.02& 15.61 $\pm$ 0.03& 14.96 $\pm$ 0.01    \\
2007.1228&16.60 $\pm$ 0.01& 16.04 $\pm$ 0.01& 15.60 $\pm$ 0.01& 14.94 $\pm$ 0.01    \\
2007.1315&16.36 $\pm$ 0.02& 15.78 $\pm$ 0.02& 15.35 $\pm$ 0.01& 14.71 $\pm$ 0.02    \\
2007.1339&16.21 $\pm$ 0.05& 15.66 $\pm$ 0.02& 15.23 $\pm$ 0.01& 14.59 $\pm$ 0.01    \\
2007.1529&      ---      & 15.63 $\pm$ 0.01& 15.18 $\pm$ 0.01& 14.53 $\pm$ 0.01    \\
2007.1750&16.19 $\pm$ 0.01& 15.64 $\pm$ 0.01& 15.21 $\pm$ 0.01& 14.56 $\pm$ 0.02    \\
2007.1777&16.21 $\pm$ 0.04& 15.62 $\pm$ 0.03& 15.20 $\pm$ 0.02& 14.59 $\pm$ 0.02    \\
2007.1802&16.30 $\pm$ 0.02& 15.72 $\pm$ 0.02& 15.28 $\pm$ 0.02& 14.64 $\pm$ 0.02    \\
2007.1939&16.20 $\pm$ 0.02& 15.66 $\pm$ 0.01& 15.22 $\pm$ 0.01& 14.60 $\pm$ 0.01    \\
2007.1967&16.16 $\pm$ 0.02& 15.61 $\pm$ 0.01& 15.20 $\pm$ 0.01& 14.58 $\pm$ 0.01    \\
2010.6989&      ---      & 16.23 $\pm$ 0.01& 15.81 $\pm$ 0.01& 15.20 $\pm$ 0.01    \\
2010.7290&      ---      & 16.09 $\pm$ 0.03& 15.64 $\pm$ 0.02& 15.04 $\pm$ 0.01    \\
2010.8059&16.01 $\pm$ 0.01& 15.55 $\pm$ 0.01& 15.12 $\pm$ 0.01&      ---  \\
2010.8085&      ---      & 15.60 $\pm$ 0.01& 15.18 $\pm$ 0.01& 14.55 $\pm$ 0.01    \\
2010.8113&      ---      & 15.63 $\pm$ 0.01& 15.22 $\pm$ 0.01& 14.59 $\pm$ 0.02    \\
2010.8278&      ---      & 15.63 $\pm$ 0.01& 15.20 $\pm$ 0.01&      ---        \\
2011.1555&15.86 $\pm$ 0.01&      ---      & 15.03 $\pm$ 0.01&      ---        \\
2017.8717&16.60 $\pm$ 0.01& 16.03 $\pm$ 0.01& 15.63 $\pm$ 0.01& 14.99 $\pm$0.01\\
2017.9182&15.82 $\pm$ 0.01& 15.26 $\pm$ 0.01& 14.89 $\pm$ 0.01& 14.29 $\pm$0.01\\
2018.0221&15.32 $\pm$ 0.01& 14.77 $\pm$ 0.01& 14.40 $\pm$ 0.01& 13.82 $\pm$0.01\\
\hline
\end{tabular}
\end{center}
\end{table*}

Older data were obtained from Perugia for the 1998-1999 fade and are presented in Table~\ref{tab:dataPer98} (for details see \cite{mas03} and \cite{cip08}). 
The data for more recent 2006, 2010 and 2017 fades are given in Table~\ref{tab:dataKRKSUH06}.
In addition to professional observatories, several amateur astronomers also contributed to the data. 
Some of these fades happen about six months after the time when the secondary is close to the jet line (expected jet-line minima at 1988.95 and 1997.61 in the binary cnetral engine model \citep{val22}) while others do not have such a connection to the orbit. 
The latest "jet-line" fade was expected around 2020.53, at the time when OJ~287 was not observable by ground-based optical telescopes \citep{val22}.

\section{Colours of OJ~287 and its host}
\label{sec:OJ_colours}
\begin{figure*}
\includegraphics[width= 0.6\textwidth]{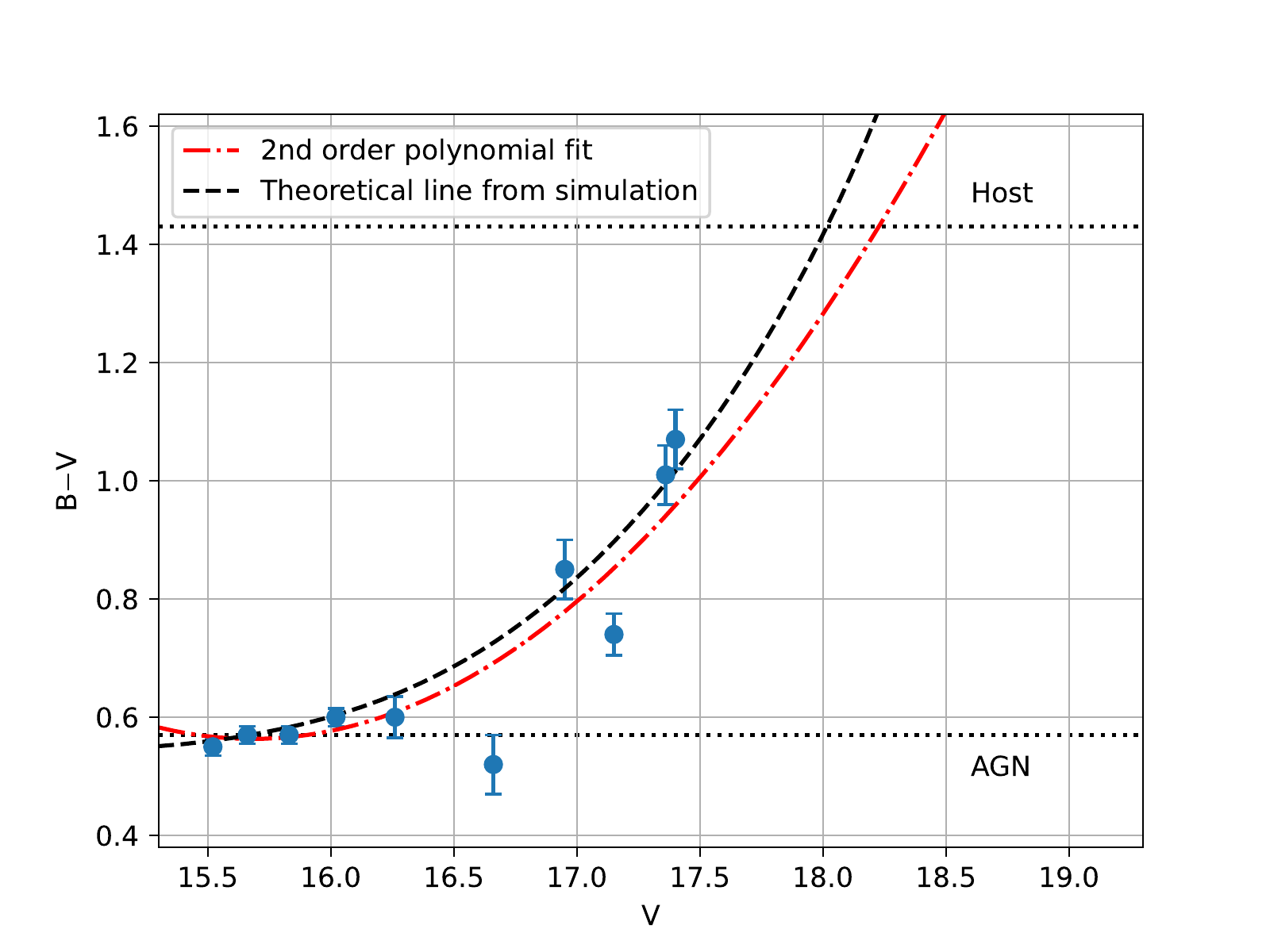}

\caption{The variation of the B-V colour as a function of V-band magnitude in OJ~287. The points above V$=$16.2 refer to the 1989 fade, the others come from our Tables~\ref{tab:dataPer98} and \ref{tab:dataKRKSUH06}, and are averages of ten measurements each. The red dash-dotted line represents a second order polynomial fit to the data points. The black dashed line shows the expected colour index variation if a host of V$=$18.0 is added to the AGN of constant colour. The horizontal dotted lines represent the expected host galaxy colour and the AGN colour, respectively.}
\label{fig:B-V_V}
\end{figure*}

In this section, we compare colours of OJ~287 at various V-magnitude levels with those corresponding to a hypothetical host galaxy. 
For a supergiant elliptical galaxy at the redshift of 0.3, the expected colours are: $B-V = 1.43 \pm 0.1$, $V-R = 0.91 \pm 0.05$, $V-I = 1.56 \pm 0.1$, and $R-I = 0.65 \pm 0.05$, for a passive evolutionary model \citep{kri78,bru83,gui87,roc88,fio97}.

For OJ~287, in addition to the steady light from the host, there is a highly variable radiation from the AGN. 
The AGN component radiates synchrotron radiation with a constant spectral index $\alpha \sim 1.5$ (the flux $F_{\nu} \sim \nu^{-\alpha}$) over the V-magnitude range from 16.5 to 14, irrespective of the flux level \citep{kid95,hag98}. 
The only confirmed exception to this rule arises during the thermal flares when the flux rise makes $V \le 13.5$ and the spectrum becomes exceptionally blue \citep{val12,val16}. 
The second possible exception, discussed in this paper, occurs when brightness in $V$ drops below about 17 mag.  
Aside from those special episodes, the colours of OJ~287 are: $B-V = 0.57 \pm 0.13$, $V-R = 0.44 \pm 0.09$ and $V-I = 1.11 \pm 0.16$ \citep{efi02}.

In order to model the gradual change of colours due to an underlying galaxy, we first carried out a second order polynomial fit to all the available colours as a function of magnitude. 
The B$-$V, V$-$R, V$-$I and R$-$I colours were taken from the values shown in Tables~\ref{tab:dataPer98} and \ref{tab:dataKRKSUH06} as well as from the previously published values in \citet{tak90}.
One such fit is shown in Figure~\ref{fig:B-V_V} for example, where we plot the B$-$V at different V intervals of OJ~287 and the red dash-dotted line shows the polynomial fit.
The expected colours of the host galaxy and the AGN are marked by the horizontal lines in these graphs (see Figure~\ref{fig:B-V_V}). 
There is an upward curvature of the best fit which is significant above $3 \sigma$ level in all except in the R-I diagram.
We identified the possible location of the host in the magnitude axis as the point where the fit to the observations meets the host line. 
At this brightness, practically all the light comes from the host galaxy.

Determined in this way, we find that the host galaxy magnitude is $V = 18.2 \pm 0.3$ from the B-V data, $V = 17.5 \pm 0.5$ from the V$-$R data, and $V = 17.25 \pm 0.5$ from the V$-$I data. 
In the R$-$I diagram, the AGN and host galaxy lines are so close to each other that we cannot derive the host magnitude. The standard deviation of the fit through points is included in the error estimate. Also the error limit includes the error in the expected colour of the host as mentioned above.

Simulations of the colours derived from the optical spectrum (power law plus host galaxy model) in different states can be used to evaluate the effective capability of the method. 
With this method we calculate the theoretical line more accurately. We add the AGN of the above mentioned colour to a host of a given V-magnitude.
Different host magnitudes were applied until we find one which gives a good fit to the observations. 
The result is shown by the black dashed line in Figure~\ref{fig:B-V_V}.
The calculations are carried out in linear flux units, and subsequently converted to magnitudes. 
The scatter in the observational points is similar to earlier studies \citep{efi02}.

From Figure~\ref{fig:B-V_V}, we find that the simulation study supports the host galaxy magnitude to be $V=18.0\pm 0.3$. 
However, the faintest R-band value, $17.0\pm 0.2$ (observed in Perugia in 1999), puts an upper limit to the host brightness of $V \sim 17.7$. 
All data are thus consistent with $V=18.0\pm0.3$.

Note that the 2017 high accuracy observations from Krakow and Mt.Suhora Observatories as well as the UK and Spanish observations, even though with a larger scatter, agree perfectly with the expected colour behaviour which arises from adding the host and the AGN components.  
This would be very surprising if the underlying reason for the colour changes is something else than the influence of the host. 
Remember that the base level of OJ~287 was about one magnitude fainter in 1989 and 1999 than in 2017. 
If the colour changes were related to the relative drop in brightness at the fade, then we should have seen a prominent upturn of 2017 points at higher V which clearly is not there.

The best spectral range for the host galaxy study is from B to V. 
This is because the expected host galaxy flux turns down strongly below the V-band wavelength while the AGN flux does not. 
In the B$-$V diagram, the six points of highest V value come from \citet{tak90}. 
The minima according to the fits to the data happen first in the R-band, half-a-day later in the V-band, and another half-a-day later in the B-band. 
If we assume that OJ~287 has the same light curve shape in all channels, but the light curves are shifted as mentioned, then the scatter in the points around the theoretical line is reduced. 
Note that the fade occurred two weeks earlier at 22 GHz and 37 GHz radio frequencies, supporting the idea of the wavelength dependence of the timing of fades.

\section{Discussion and conclusions}
\label{sec:discussion}
There have been several attempts to measure the host galaxy magnitude by direct imaging, after subtracting the point-like AGN from the image. 
The task is difficult since the brightness difference between the host and the AGN is typically about 2 magnitudes in the K-band, 3 magnitudes in the I-band, and even greater at shorter optical wavelengths \citep{nil19}.
The most favourable contrast between the AGN and the host so far has been obtained in infrared. 
\citet{nil19} reported recent infrared observations at the 2.5-meter \emph{Nordic Optical Telescope} when the AGN was at a low state. 
Imaging data were also obtained at the 10.4-meter \emph{Gran Telescopio Canarias} somewhat after the 2017 deep fade. 
They essentially confirm the findings of this paper, as we will discuss below.

We have determined the magnitude of the host galaxy of the BL Lacertae object OJ~287 using the change of the colour of the object during its minimum light. 
It is interesting to translate the result to the K-band for a direct comparison with the surface brightness method. 
Using the distance modulus m - M = 41, V - K = 3.2 \citep{fio97,ho11}, and the combined K-correction and evolution correction of -0.3 in the K-band \citep{kri78,bru83,gui87,roc88,fio97,chi10,nil19}, we obtain the absolute K-band magnitude of OJ~287 $M_{K}=-26.5\pm0.3$. 
This may be compared with the absolute K-band magnitude $M_{K}=-26.8$ of the nearby elliptical galaxy NGC~4889; the other nearby supergiant NGC~1600 is a little fainter \citep{tho16}. 
These two galaxies are interesting as their central black hole masses agree with the primary black hole mass in OJ~287 within error limits \citep{val10}. 

\begin{figure}
\includegraphics[width=0.45\textwidth]{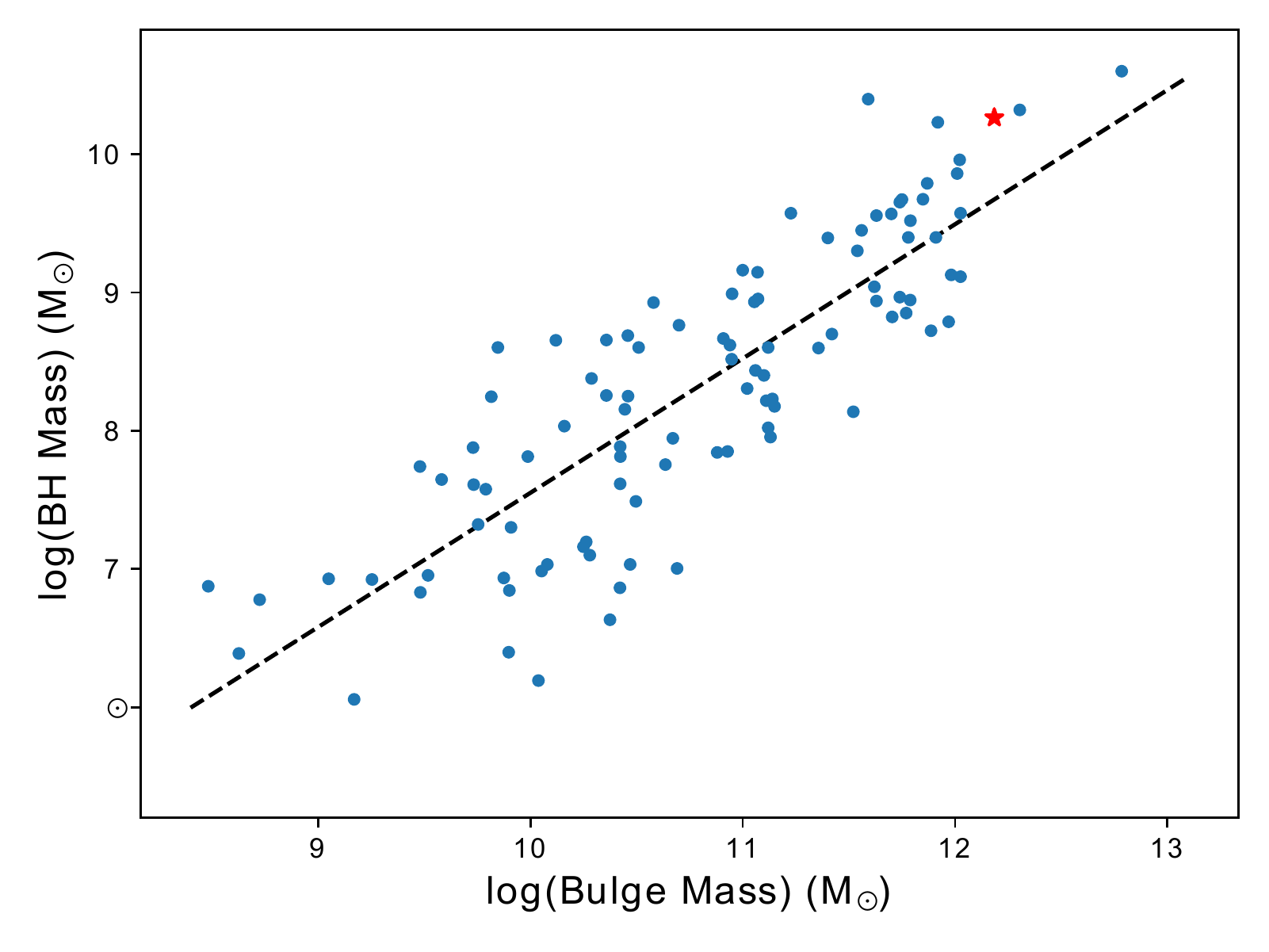}
\caption{The galaxy sample of \citet{sag16}
in the black hole mass vs. galaxy stellar mass diagram (blue points), while the position of OJ~287 is given as a red star. OJ~287 is taken to be $0.3$ magnitudes fainter than NGC~4889. Three recent additional measurements have been added: NGC~1600 \citep{tho16}, Abell 1201 central galaxy \citep{smi17} and Holm 15A \citep{meh19}. We note that OJ~287 lies within the one sigma scatter from the mean linear correlation (dashed line).} 
\label{fig:BHmass-Bulgemass}
\end{figure}

In Figure~\ref{fig:BHmass-Bulgemass}, we place OJ~287 in the BH mass vs. galaxy mass diagram \citep{sag16}.
We see from Figure~\ref{fig:BHmass-Bulgemass} that OJ~287 is slightly, but not significantly, above the line for the mean linear correlation for nearby galaxies. 
If it were placed in the corresponding graph for galaxies at redshift 0.5 \citep{por12}, the result would be very much the same.

All studies of the magnitude of OJ~287 by the imaging method have been limited to the distance range 5 - 35 kpc from the center.
The light from the rest of the galaxy has to be extrapolated using models. 
\cite{nil19} find the K-band magnitudes, K=14.25 and K=15.27, where the first one has the higher signal-to-noise ratio ($\sim 8$) of the two. 
This translates to $M_{K}=-26.45\pm0.1$ with the same K-correction and evolution correction as above. 
The second K-band determination would make the host one magnitude fainter. 
Either way, the host is found to be very bright also by using the surface brightness method in the K-band.

Experience with other elliptical galaxies in this brightness class (e.g. M87) has shown that if we restrict our measurements within 30 kpc distance from the center, we are missing more than a negligible amount of light from the outer envelope of the galaxy.
For example, the absolute V-magnitude of NGC~4889 \citep{san72} is $M_{V}=-22.2$ (using m - M = 34.7) measured out to roughly the same 30 kpc linear distance as the K-band data of \citet{nil19}, while its total magnitude out to 100 kpc distance is about $M_{V}=-23.6$ \citep{gra13}. 
Typically the outer envelope contributes about 0.8 magnitudes to the total brightness \citep{dev70,las14,hua13}.

The colour method measures all the light from the galaxy, including the outer envelope within the diaphragm of the photometer. 
In this case, it is roughly equivalent to 35 kpc linear distance at OJ~287. 
Therefore some of the light of the host galaxy is missed by this method. 
In the imaging method the outer envelope may be missed altogether if the data are not sufficient to include it in the model. 
In OJ~287, the point spread function of the central point source prevents the determination of the accurate sky background, necessary to measure the magnitude of the outermost component. 
Going towards shorter wavelengths, the signal-to-noise ratio becomes considerably worse due the AGN component becoming more and more dominant. 
In the I-band the signal-to-noise ratio falls below 5 \citep{nil19}, and at shorter wavelengths the situation becomes even worse.
Probably for this reason, some other surface brightness measurements seem to have seriously underestimated the true brightness of the OJ~287 host galaxy \citep{wur96,hei99}. 
The Hubble Space Telescope I-band measurement by \cite{yan97}, translated to K-band, gives the brightness upper limit at $M_{K}\sim-26.3$, in agreement with our result especially if the missing low brightness outer halo is taken into account.

Our magnitude determination is based on the assumption that the colour of OJ~287 at low light level is modified by the light of the host galaxy. 
It is also possible that there is a "cosmic conspiracy", that is, the colour of OJ~287 evolves at low light levels because the AGN mimics the effects of the host. 
However, in this case, one might assume that the colour behaviour should be related to the relative flux decline, and we should see a similar behaviour in 2017 as in 1989 and 1999. 
On the other hand, if the colour changes are related to the absolute brightness levels, such as if the constant host galaxy is influencing the colours, then the 2017 fade should be quite different from the earlier deep fades. 
The latter has clearly been the case.

It appears that the lowering the AGN activity does not in itself cause the observed colour changes. 
If indeed the base level brightness variation is due to the varying Doppler boosting \citep{val13}, then the current 2017 fade is about as deep as the 1989-1999 fades in relative terms. 
The colour changes should be seen now as they were seen on earlier occasions. 
The three high accuracy points from Krakow and Mt.Suhora Observatories during the last fade (see the the first three points on the left hand side of Figure~\ref{fig:B-V_V}) demonstrate that this is not the case.

\cite{kom20,kom21} confirm this result over a wider spectral range, using SEDs from the optical to UV (and X-rays as well), taken with the Neil Gehrels Swift observatory (Swift hereafter) at a cadence of 1--2 days in the course of the program MOMO (Multiwavelength observations and modelling of OJ 287; see \citet{kom21b}).
All optical and UV bands, and the X-rays are near simultaneous, the UV-optical bands within minutes. 
In particular, \citet{kom21} reported that the ratio of UV flux over optical flux ratio is constant throughout the deep fade, and therefore excluded an extinction event, because the Swift UV bands are very sensitive to extinction. 
Table~\ref{tab:dataAMAT} gives the average B-V colour and the V-band over UV-band W2 flux ratio at 4 brightness levels.
It shows no significant colour trends with brightness. 
Going to the other direction on the wavelength axis, \cite{kid98} have found that between infrared and optical regions, there are no colour changes over the V-magnitude range from 14.7 to 15.7.

\begin{table}
\begin{center}
\caption{Summary of Swift observations.}
\label{tab:dataAMAT}
\begin{tabular}{lcc}
\hline
$V$&$B-V$&  $V/W2$    \\
\hline
  16.0&  0.55  $\pm$ 0.03& 1.47 $\pm$ 0.05\\
  15.6&  0.60  $\pm$ 0.03& 1.58 $\pm$ 0.05\\
  15.2&  0.54  $\pm$ 0.03& 1.49 $\pm$ 0.05\\
  14.8&  0.53  $\pm$ 0.03& 1.46 $\pm$ 0.05\\
\hline
\end{tabular}
\end{center}
\end{table}

The colour changes are apparently related to the absolute brightness of OJ~287. The influence of the host galaxy naturally explains this fact and seems to exclude the possibility that there is something in the jet emission mechanism that changes the colour depending on the blazar flux level.

It is likely that deep fades will happen in OJ~287 also in future. 
They will provide an excellent opportunity to probe the outer envelope of this galaxy further. 
There is no reason why the photometry during the fade should be limited only within 6 arcsec from the center.
While the direct imaging of the outer envelope is challenging, the separation of the host from the AGN by the colour method is rather simple. 
And on the basis of the excellent agreement of the magnitudes of the host galaxy by the two different methods we have demonstrated, the colour separation is a reliable method, and it could be recommended for the study of host galaxies in other AGN as well. 
The method is particularly well suited for the detection of low brightness level components of the host galaxies.
Suitable candidates for such wide aperture photometry are found among the lowest redshift BL Lacertae objects \citep{urr00}. 
To our knowledge, such a study has never been attempted.
It is understandable, since it takes a large amount of observing time to sample the BL Lacertae objects at different light levels, and moreover, the result cannot be deduced from usual photometry where the aperture is chosen so as to eliminate the influence of the host on the AGN magnitude.

The fades may arise from a temporary misalignment of the jet with regard to its average direction \citep{tak90}. 
The jet appears to wiggle \citep{gom21} and in principle these wiggles are predictable \citep{dey21}. 
The modelling of the time behaviour of the optical polarisation may be a way to predict also the optical fades. 
Alternatively, the fades may represent times of a change in accretion flow affecting (reducing) the jet emission, and it may also be possible to predict such moments of time \citep{val22}.

\section*{Acknowledgements}

We would like to thank Roberto Saglia for providing information on black hole mass measurements in advance of publication and for enlightening discussions. S.Z. would also like to acknowledge support of the NCN grant No. 2018/29/B/ST9/01793, and K.M. JSPS KAKENHI grant number 19K03930.

\section*{Data Availability}

The data published in this paper is available upon request from the authors.











\bsp	
\label{lastpage}
\end{document}